\documentclass{amsart}

\usepackage{amssymb,xspace}

\theoremstyle{definition}

\theoremstyle{remark}

\numberwithin{equation}{section}

\newcommand{\ket}[1]{\ensuremath{|#1\rangle}\xspace}

\begin{document}

\title[BI polynomials as Racah coefficients of $sl_{-1}(2)$]{The Bannai-Ito polynomials as Racah coefficients of the $sl_{-1}(2)$ algebra}


\author{Vincent X. Genest}
\address{Centre de recherches math\'ematiques, Universit\'e de Montr\'eal, C.P. 6128, Succursale Centre-ville, Montr\'eal, Qu\'ebec, Canada, H3C 3J7}
\email{genestvi@crm.umontreal.ca}
\author{Luc Vinet}
\address{Centre de recherches math\'ematiques, Universit\'e de Montr\'eal, C.P. 6128, Succursale Centre-ville, Montr\'eal, Qu\'ebec, Canada, H3C 3J7}
\email{luc.vinet@umontreal.ca}
\author{Alexei Zhedanov}
\address{Donetsk Institute for Physics and Technology, Donetsk 83114, Ukraine }
\email{zhedanov@yahoo.com}

\date{}

\dedicatory{}

\commby{}

\begin{abstract}
The Bannai-Ito polynomials are shown to arise as Racah coefficients for $sl_{-1}(2)$. This Hopf algebra has four generators including an involution and is defined with both commutation and anticommutation relations. It is also equivalent to the parabosonic oscillator algebra. The coproduct is used to show that the Bannai-Ito algebra acts as the hidden symmetry algebra of the Racah problem for $sl_{-1}(2)$. The Racah coefficients are recovered from a related Leonard pair.
\end{abstract}

\maketitle

\section*{Introduction}
\normalsize
\thispagestyle{empty}
The $sl_{-1}(2)$ algebra was introduced recently in \cite{Vinet-2011} as a deformation of the classical $sl(2)$ Lie algebra; it is defined in terms of four generators, including an involution, satisfying both commutation and anticommutation relations. This algebra can also be obtained from the quantum algebra $sl_{q}(2)$ by taking the limit $q\rightarrow -1$ and is furthermore the dynamical algebra of a parabosonic oscillator \cite{Green-1953,Mukunda-1980}. We here consider the Racah problem for this algebra.

Recently, a series of orthogonal polynomials corresponding to limits $q\rightarrow -1$ of $q$-polynomials of the Askey scheme were discovered \cite{VZhedanov-2011,Vinet-2012,Vinet-2011-2,Zhedanov-2011}. These polynomials are eigenfunctions of operators of Dunkl type, which involve the reflection operator \cite{Dunkl-1998,Vinet-2011-3}. Interestingly, these polynomials have also been related to Jordan anticommutator algebras \cite{Tsujimoto-2011}. In most references, so far, these $q=-1$ polynomials have been left buried in the standard classifications. In view of their bispectrality and remarkable properties, a $-1$ scheme would deserve to be highlighted. 

At the top of the discrete variable branch of this $q=-1$ class of polynomials lie the Bannai-Ito (BI) polynomials \cite{Ito-1984} and their kernel partners, the complementary Bannai-Ito polynomials \cite{Vinet-2012}. Both sets depend on four parameters and are expressible in terms of Wilson polynomials \cite{Ito-1984,Koekoek-2010,Vinet-2012}. The BI polynomials possess the Leonard duality property, which in fact led to their initial discovery in \cite{Ito-1984}. In contradistinction, the complementary BI polynomials and their descendants, the dual $q=-1$ Hahn polynomials \cite{VZhedanov-2011}, are also bispectral but fall beyond the scope of the Leonard duality.

 The Clebsch-Gordan problem for $sl_{-1}(2)$ was first solved in \cite{Vinet-2011}; it was shown that the coupling coefficients for two $sl_{-1}(2)$ algebras, also called Clebsch-Gordan or Wigner coefficients, are proportional to the dual $q=-1$ Hahn polynomials \cite{VZhedanov-2011}. In this paper, we investigate the Racah problem for $sl_{-1}(2)$, which is tantamount to finding the coupling coefficients for three parabosonic oscillators. It is shown that these coefficients are also expressed in terms of $q=-1$ polynomials, in this case the Bannai-Ito polynomials. Our approach consists in constructing the Jordan algebra of the intermediary Casimir operators that appear in the coproduct \cite{Daska-2000} of three $sl_{-1}(2)$ algebras; this anticommutator algebra coincides with the Bannai-Ito algebra \cite{Vinet-2012}, a special case of the Askey-Wilson algebra introduced in \cite{Zhedanov-1991}. The two Casimir operators are then shown to form a Leonard pair \cite{Brown-2012,Curtin-2007,Huang-2011,Terwilliger-2001,Terwilliger-1999,Terwilliger-2008}, an observation which allows to recover the recurrence relation of the Bannai-Ito polynomials for the overlap (Racah) coefficients.

The outline of the paper is as follows. In section 1, we recall the definition of the $sl_{-1}(2)$ algebra, its irreducible representations and its coproduct structure. We also provide a review of the theory of the Bannai-Ito polynomials and go over the basics of Leonard pairs and the corresponding Askey-Wilson relations \cite{Terwilliger-2001,Vidunas-2008}. In section 2, we review the Clebsch-Gordan problem for the parabosonic algebra $sl_{-1}(2)$. In section 3, we show that the intermediary Casimir operators $(K_1,K_2)$ of the sum of three $sl_{-1}(2)$ algebras form the Bannai-Ito algebra. In section 4, the operators $(K_1,K_2)$ are re-expressed as a Leonard pair which is used to recover the recurrence relation satisfied by the overlap coefficients (Racah) coefficients. The exact expression for the Racah coefficients is finally obtained up to a phase factor using the orthogonality relation of the BI polynomials. In section 5, we discuss the degenerate case of the Bannai-Ito algebra corresponding to the anticommutator spin algebra \cite{Gorodnii-1984, Silvestrov-1992,Arik-2003,Brown-2012}. We conclude by explaining that the operators $K_1$ and $K_2$, together with their anticommutator $K_3$, form a Leonard triple. A different Racah problem, which involves modifying the addition rule of $sl_{-1}(2)$, is considered to that end.

\section{The $sl_{-1}(2)$ algebra, Bannai-Ito polynomials and Leonard pairs}
\subsection{$sl_{-1}(2)$ essentials}
The Hopf algebra $sl_{-1}(2)$ \cite{Vinet-2011} is generated by four operators $J_{0}$, $J_{+}$, $J_{-}$ and $R$ satisfying the relations
\begin{align}
\label{Definition}
[J_0,J_{\pm}]=\pm J_{\pm}, && [J_0,R]=0, && \{J_{+},J_{-}\}=2J_0, &&& \{J_{\pm}, R\}=0,
\end{align}
where $[x,y]=xy-yx$ and $\{x,y\}=xy+yx$. The operator $R$ is an involution, which means that it satisfies the property
$$
R^2=\mathfrak{id},
$$
where $\mathfrak{id}$ is the identity. The Casimir operator, which commutes with all $sl_{-1}(2)$ elements, is given by
\begin{equation}
\label{Casimir}
\mathcal{Q}=J_{+}J_{-}R-J_0R+R/2.
\end{equation}
Let $\epsilon=\pm 1$ and $\mu\geqslant0$ be two parameters; we denote by $(\epsilon,\mu)$ the infinite-dimensional vector space spanned by the basis $\ket{\epsilon;\mu;n}$, $n\in \mathbb{N}$, endowed with the actions
\begin{align}
\label{action}
\begin{split}
J_{0}\ket{\epsilon;\mu;n}&=(n+\mu+1/2)\ket{\epsilon;\mu;n}, 
\;\;\;\;\;R\ket{\epsilon;\mu;n}=\epsilon (-1)^{n}\ket{\epsilon;\mu;n},\\
J_{+}\ket{\epsilon;\mu;n}&=\sqrt{[n+1]_{\mu}}\ket{\epsilon;\mu;n+1}, 
\;\;\;\;\;J_{-}\ket{\epsilon;\mu;n}=\sqrt{[n]_{\mu}}\ket{\epsilon;\mu;n-1},
\end{split}
\end{align}
where $[n]_{\mu}$ denotes the $\mu$-number
\begin{align}
[n]_{\mu}=n+\mu(1-(-1)^{n}).
\end{align}
With the actions \eqref{action}, the vector space $(\epsilon,\mu)$ forms an irreducible $sl_{-1}{(2)}$-module. On this module, the Casimir operator is a multiple of the identity 
\begin{align}
\label{Casimir-action}
\mathcal{Q}\ket{\epsilon;\mu;n}=-\epsilon\,\mu\ket{\epsilon;\mu;n},
\end{align}
as expected from Schur's lemma. On the space $(\epsilon,\mu)$, the algebra $sl_{-1}(2)$ is equivalent to the parabosonic oscillator algebra. Indeed, one has
$$
[J_{-},J_{+}]=\{J_{-},J_{+}\}-2J_{+}J_{-}=2J_{0}-2J_{+}J_{-}.
$$
Using the expression \eqref{Casimir} for the Casimir operator and its action on vectors of $(\epsilon,\mu)$, we find
\begin{equation}
\label{Para}
[J_{-},J_{+}]=1+2\epsilon\mu R.
\end{equation}
The operators $J_{\pm}$ satisfying the commutation relation \eqref{Para}, together with the operator $R$ obeying  the relations $R^2=\mathfrak{id}$ and $\{R,J_{\pm}\}=0$, define the parabosonic oscillator algebra \cite{Daska-2000,Mukunda-1980,Rosenblum-1994}.
 
The algebra $sl_{-1}(2)$ admits a non-trivial addition rule, or coproduct. Let $(\epsilon_1,\mu_1)$ and $(\epsilon_2,\mu_2)$ be two $sl_{-1}(2)$-modules. A third module can be obtained by taking tensor product $(\epsilon_1,\mu_1)\otimes(\epsilon_2,\mu_2)$ equipped with the transformations
\begin{align}
\begin{split}
\label{action-co}
J_0(v\otimes w)&=(J_0 v)\otimes w + v\otimes(J_0w),\\
J_{\pm}(v\otimes w)&=(J_{\pm}v)\otimes (Rw)+v\otimes(J_{\pm}w),\\
R(v\otimes w)&=(Rv)\otimes(Rw),
\end{split}
\end{align}
where $v\in (\epsilon_1,\mu_1)$ and $w\in(\epsilon_2,\mu_2)$. The addition rule for $sl_{-1}(2)$ can also be presented without referring to any representation. Let $J_0^{(i)}$, $J_{\pm}^{(i)}$ and $R^{(i)}$ be two  mutually commuting sets of $sl_{-1}(2)$ generators. A third algebra, denoted symbolically $3=1\oplus 2$, is obtained by defining
\begin{align}
\label{addition-rule}
J_0^{(3)}=J_0^{(1)}+J_{0}^{(2)}, && J_{\pm}^{(3)}=J_{\pm}^{(1)}R^{(2)}+J_{\pm}^{(2)}, && R^{(3)}=R^{(1)}R^{(2)}.
\end{align}
It is easily verified that the generators $J_{0}^{(3)}$, $J_{\pm}^{(3)}$ and $R^{(3)}$ satisfy the defining relations of $sl_{-1}(2)$ given in \eqref{Definition}. The Casimir operator for the third algebra, denoted by $\mathcal{Q}_{12}$, is
\begin{equation}
\label{Casimir-added}
\mathcal{Q}_{12}=J_{+}^{(3)}J^{(3)}_{-}R^{(3)}-J_0^{(3)}R^{(3)}+(1/2)\,R^{(3)}
\end{equation}

\subsection{Bannai-Ito polynomials}
Bannai and Ito discovered their polynomials in 1984 in their complete classification of orthogonal polynomials satisfying the Leo\-nard duality property \cite{Ito-1984}. These polynomials were shown to be $q=-1$ limits of the $q$-Racah polynomials and many of their properties (e.g. recurrence relation, weight function, hypergeometric representation) were given in their book \cite{Ito-1984}. Recently, it was shown in \cite{Vinet-2012} that the Bannai-Ito polynomials also occur naturally as eigensolutions of Dunkl shift operators.In the following, we review some of the properties of the BI polynomials. 

The monic BI polynomials satisfy the recurrence relation
\begin{equation}
\label{Recurrence-BI}
P_{n+1}(x)+(\rho_1-A_n-C_n)P_{n}(x)+A_{n-1}C_nP_{n-1}(x)=xP_{n}(x),
\end{equation}
where 
\begin{align}
\label{Coeff-1}
A_{n}&=
\begin{cases}
\frac{(n+1+2\rho_1-2r_1)(n+1+2\rho_1-2r_2)}{4(n+1-r_1-r_2+\rho_1+\rho_2)}, & \text{$n$ even},\\
\frac{(n+1-2r_1-2r_2+2\rho_1+2\rho_2)(n+1+2\rho_1+2\rho_2)}{4(n+1-r_1-r_2+\rho_1+\rho_2)}, & \text{$n$ odd},
\end{cases}
\\
C_{n}&=
\begin{cases}
-\frac{n(n-2r_1-2r_2)}{4(n-r_1-r_2+\rho_1+\rho_2)}, & \text{$n$ even},\\
-\frac{(n-2r_2+2\rho_2)(n-2r_1+2\rho_2)}{4(n-r_1-r_2+\rho_1+\rho_2)}, & \text{$n$ odd}.
\end{cases}
\end{align}
The polynomials satisfying \eqref{Recurrence-BI} are called positive definite if $U_n=A_{n-1}C_n>0$ for all $n\geqslant 1$. This condition is also equivalent to the existence of a positive orthogonality measure for the polynomials $P_{n}(x)$. In the case of the BI polynomials, it is seen that this condition cannot be fulfilled for all values of $n$. However, if $U_i>0$ for $i=1,\ldots,N$ and $U_{N+1}=0$, it is known that one has a finite system of orthogonal polynomials $P_0(x),\,P_1(x),\ldots,\,P_N(x)$ satisfying the discrete orthogonality relation
\begin{equation}
\label{Orthogonality}
\sum_{s=0}^{N}\omega_s(x_s)P_{n}(x_s)P_m(x_s)=h_n\delta_{nm},\;\;\;\;\;\;h_n=u_1,\ldots,u_n,
\end{equation}
on the lattice $x_s$, where $s=0,1,\ldots,N$. The discrete points $x_s$ are the simple roots of the polynomial $P_{N+1}(x)$ \cite{Chihara-1978}.

When $N$ is an even integer, the truncation condition $U_{N+1}=0$ is equivalent to one of the four possible conditions 
\begin{equation}
\label{conditions}
2(r_i-\rho_k)=N+1,\;\;\;\;\;i,k=1,2.
\end{equation}
The case of relevance here is
\begin{equation}
2(r_2-\rho_1)=N+1.
\end{equation}
We introduce the following parametrization:
\begin{align}
\label{Para-Pair}
\begin{split}
2\rho_1=(b+c),\;\;\;\;\; & 2\rho_2=(2a+b+c+N+1),\\
 2r_1=(c-b),\;\;\;\;\; & 2r_2=(b+c+N+1),
\end{split}
\end{align}
where $a$, $b$ and $c$ are arbitrary positive parameters. Assuming \eqref{Para-Pair}, the coefficient $U_n$ takes the form:
\begin{equation}
U_n=
\begin{cases}
\frac{n(N+2c+1-n)(n+2a+2b)(n+2a+2b+2c+N+1)}{16(a+b+n)^2}, & \text{$n$ even,}\\
\frac{(N+1-n)(2a+n)(2b+n)(n+2a+2b+N+1)}{16(a+b+n)^2}, & \text{$n$ odd.}
\end{cases}
\end{equation}
From this expression, it is obvious that $U_{N+1}=0$ and that the positivity condition $U_n>0$ is satisfied for  $n=0,\ldots,N$. With this parametrization, the Bannai-Ito polynomials obey the orthogonality relation
\begin{equation}
\label{Ortho-Pair}
\textstyle
\sum_{\ell=0}^{N}\Omega_{\ell}P_{n}(x_{\ell})P_{m}(x_{\ell})=\Phi_{N,n}\delta_{nm}.
\end{equation}
The orthogonality grid is given by
\begin{equation}
\label{Grille-1}
\textstyle
x_{\ell}=\frac{1}{2}\left[(-1)^{\ell}(\ell+b+c+1/2)-1/2\right].
\end{equation}
The weight function $\Omega_{\ell}$ takes the form
\begin{equation}
\label{weight-even}
\Omega_{\ell}=(-1)^{q}\frac{(-N/2)_{k+q}(1/2+b)_{k+q}(1+b+c)_{k}(3/2+a+b+c+N/2)_{k}}{(1/2+c)_{k+q}(1+b+c+N/2)_{k+q}(1/2-a-N/2)_{k}k!},
\end{equation}
where $\ell=2k+q$ with $q=0,1$ and where $(x)_n=(x)(x+1)\cdots (x+n-1)$ stands for the Pochhammer symbol. Furthermore, the normalization factor $\Phi_{N,n}$ is found to be
\begin{multline}
\Phi_{N,n}=\frac{m!k!}{(m-k-q)!}\left[\frac{(1+a+b+k)_{m-k}(1+b+c)_{m}}{(1/2+a+k+q)_{m-k-q}(1/2+c)_{m-k}}\right]\\
\times \left[\frac{(1/2+b)_{k+q}(m+1+a+b)_{k+q}(m+3/2+a+b+c)_{k}}{(k+1+a+b)_{k+q}^2}\right],
\end{multline}
where $m=N/2$ and $n=2k+q$ with $q=0,1$. The other truncation conditions in \eqref{conditions} can be treated similarly.

When $N$ is an odd integer, the truncation condition $U_{N+1}=0$ is equivalent to one of the three conditions
\begin{align}
\begin{split}
i)\;\rho_1&+\rho_2=-\frac{N+1}{2},\;\;\;\;\;\;\; ii)\;r_1+r_2=\frac{N+1}{2},\\
& iii)\;\rho_1+\rho_2-r_1-r_2=-\frac{N+1}{2}.
\end{split}
\end{align}
The condition $iii)$ leads to a singular $U_n$ for $n=(N+1)/2$. Consequently, only the conditions  $i)$ and $ii)$ are admissible. The case of relevance here is
\begin{equation}
2(\rho_1+\rho_2)=-(N+1).
\end{equation}
We introduce the following parametrization:
\begin{align}
\label{Para-Impair}
\begin{split}
2\rho_1&=(\beta+\gamma),\;\;\;\;\; 2\rho_2=-(\beta+\gamma+N+1),\\
2r_1&=(\gamma-\beta),\;\;\;\;\; 2r_2=-(2\alpha+\beta+\gamma+N+1),
\end{split}
\end{align}
where $\alpha$, $\beta$ and $\gamma$ are arbitrary positive parameters. Assuming \eqref{Para-Impair}, the coefficient $U_n$ becomes
\begin{equation}
U_{n}=
\begin{cases}
\frac{n(N+1-n)(n+2\alpha+2\beta)(n+2\alpha+2\beta+N+1)}{16(\alpha+\beta+n)^2}, & \text{$n$ even,}\\
\frac{(N+2\gamma+1-n)(2\alpha+n)(2\beta+n)(n+2\alpha+2\beta+2\gamma+N+1)}{16(\alpha+\beta+n)^2}, & \text{$n$ odd.}
\end{cases}
\end{equation}
In this form, the truncation and positivity conditions are manifestly satisfied. With these parameters, the Bannai-Ito polynomials obey the orthogonality relation
\begin{equation}
\label{Ortho-Odd}
\sum_{\ell=0}^{N} \Omega_{\ell}P_{n}(x_{\ell})P_{m}(x_{\ell})=\Phi_{N,n}\delta_{nm}.
\end{equation}
The grid is given by
\begin{equation}
\label{Grille-2}
x_{\ell}=\frac{1}{2}\left[(-1)^{\ell}(\ell+\beta+\gamma+1/2)-1/2\right].
\end{equation}
The weight function takes the form
\begin{align}
\label{weight-odd}
\Omega_{\ell}=(-1)^{q}\frac{(\frac{1-N}{2})_{k}(\frac{1}{2}+\beta)_{k+q}(1+\beta+\gamma)_{k}(1+\alpha+\beta+\gamma+\frac{N}{2})_{k+q}}{(\frac{1}{2}+\gamma)_{k+q}(-\alpha-\frac{N}{2})_{k+q}(\frac{3}{2}+\beta+\gamma+\frac{N}{2})_kk!},
\end{align}
where $\ell=2k+q$ with $q=0,1$ and the normalization factor can be evaluated to
\begin{multline}
\Phi_{N,n}=\frac{(m-1)!k!}{(m-k-1)!}\left[\frac{(1+\alpha+\beta+k)_{m-k}(1+\beta+\gamma)_m}{(1/2+k+q+\alpha)_{m-k-q}(1/2+\gamma)_{m-k-q}}\right]\\
\times\left[\frac{(1/2+\beta)_{k+q}(m+1+\alpha+\beta)_k(m+1/2+\alpha+\beta+\gamma)_{k+q}}{(k+1+\alpha+\beta)_{k+q}^2}\right],
\end{multline}
where $m=(N+1)/2$ and $n=2k+q$ with $q=0,1$.

The Bannai-Ito polynomials correspond to the limit $q\rightarrow -1$ of the classical Wilson polynomials and admit a hypergeometric representation. The truncated generalized hypergeometric series is defined by
\begin{equation}
{}_{p+1}F_q\left(\begin{matrix}-n,\;a_1,\ldots,a_p\\ \phantom{-}b_1,\;b_2,\ldots,b_q\;\end{matrix};x\right)=\sum_{j=0}^{n}\frac{(-n)_j(a_1)_j\cdots(a_p)_j}{(b_1)_j(b_2)_j\cdots(b_q)_j}\frac{x^j}{j!}.
\end{equation}
We define
\begin{align}
W_{2n}(x)&=\kappa_{n}^{(1)}\, {}_4F_3\left(
\begin{matrix}
-n,\;  n+g+1,\;  \rho_2+x,\;  \rho_2-x\\
\rho_1+\rho_2+1,\;  \rho_2-r_1+\frac{1}{2},\;   \rho_2-r_2+\frac{1}{2} 
\end{matrix};1
\right),\\
W_{2n+1}(x)&=\kappa_{n}^{(2)}(x-\rho_2)\,{}_4F_3\left(
\begin{matrix}
-n,\;  n+g+2,\;  \rho_2+1+x,\;  \rho_2+1-x\\
\rho_1+\rho_2+2,\;  \rho_2-r_1+\frac{3}{2},\;   \rho_2-r_2+\frac{3}{2} 
\end{matrix};1
\right),
\end{align}
with $g=\rho_1+\rho_2-r_1-r_2$ and where the factors which ensure that the polynomials are monic are given by
\begin{align}
\label{dompe}
\kappa_n^{(1)}&=\frac{(1+\rho_1+\rho_2)_n(\rho_2-r_1+1/2)_n(\rho_2-r_2+1/2)_n}{(n+g+1)_n},\\
\kappa_n^{(2)}&=\frac{(2+\rho_1+\rho_2)_n(\rho_2-r_1+3/2)_n(\rho_2-r_2+3/2)_n}{(n+g+2)_n}.
\end{align}
The monic BI polynomials have the following expression:
\begin{align}
\label{Hypergeo-representation}
P_n(x)=W_{n}(x)-C_nW_{n-1}(x),
\end{align}
where $C_n$ is given by \eqref{Coeff-1}.
\subsection{Leonard pairs and Askey-Wilson relations}
Let $V$ be a $\mathbb{C}$-vector space of dimension $N+1$. A square matrix $X$ is said \emph{irreducible tridiagonal} if each of its non-zero entry lies on either the diagonal, sub-diagonal or super-diagonal and if each entry on the super-diagonal and sub-diagonal are non-zero. A \emph{Leonard pair} on $V$ is an ordered pair of linear transformations $(K_1,K_2)\in \mathrm{End}\,{V}$ satisfying the following conditions \cite{Terwilliger-2001}:
\begin{itemize}
\item There exists a basis for $V$ with respect to which the matrix representing $K_1$ is diagonal and the matrix representing $K_2$ is irreducible tridiagonal.
\item There exists a basis for $V$ with respect to which the matrix representing $K_2$ is diagonal and the matrix representing $K_1$ is irreducible tridiagonal.
\end{itemize}
 Leonard pairs have deep connections with orthogonal polynomials on finite grids and have also appeared in combinatorics \cite{Ito-1984,Terwilliger-2001}. Given a Leonard pair $(K_1,K_2)$, it is known \cite{Terwilliger-2008,Vidunas-2008,Zhedanov-1991} that $K_1$, $K_2$ obey the so-called Askey-Wilson relations
\begin{align}
\label{AW-relation}
\begin{split}
&K_1^2K_2-\beta K_1K_2K_1+K_2K_1^2-\gamma_1\{K_1,K_2\}-\rho_1 K_2=\gamma_2K_1^2+\omega K_1+\eta_1 \mathfrak{id},\\
&K_2^2K_1-\beta K_2K_1K_2+K_1K_2^2-\gamma_2\{K_1K_2\}-\rho_2K_2=\gamma_1K_1^2+\omega K_1+\eta_2 \mathfrak{id},
\end{split}
\end{align}
with scalars $\{\beta,\gamma_{i},\eta_i,\rho_i\}\in\mathbb{C}$. These scalars are uniquely defined provided that the dimension of the vector space is at least $4$. The converse is not always true. Indeed, if one sets $\beta=q+q^{-1}$ and $q$ a root of unity, two linear transformations obeying relations \eqref{AW-relation} do not necessarily form a Leonard pair \cite{Terwilliger-1999}. In the present work we will nonetheless obtain a Leonard pair satisfying the relations \eqref{AW-relation} with $q=-1$.

We briefly recall  how orthogonal polynomials occur in this context. Consider a Leonard pair $(K_1,K_2)$ on a vector space $V$ of dimension $N+1$. By definition, the eigenvalues of $K_1$ and $K_2$ are mutually distinct. Denoting the eigenvalues of $K_1$ by $\lambda_i^{(1)}$ for $i=0,1,\ldots,N$, there exists a basis of $V$ in which the matrices representing $K_1$ and $K_2$ are of the form
\begin{align}
\label{Matrices}
K_1=\begin{pmatrix}
\lambda_0^{(1)} &  &    & \mathbf{0} \\
 & \lambda_1^{(1)} &    &  \\
 &  & \ddots  &  \\
\mathbf{0} &  &  & \lambda_N^{(1)} 
\end{pmatrix}, &&
K_2=\begin{pmatrix}
a_0 & c_1 &  &  & \mathbf{0} \\
x_0 & a_1 & c_2 &  &  \\
 & x_1 & a_2 & \ddots & \\
 &  & \ddots & \ddots &c_{N}\\
\mathbf{0} &  &  & x_{N-1} & a_N
\end{pmatrix} .
\end{align}
One can define the sequence of  polynomials $p_i$ with $i=0,\ldots,N$ and initial condition $p_{-1}=0$ satisfying the recurrence relation
\begin{equation}
\label{recurrence-gene}
y\,p_i(y)=c_{i+1}p_{i+1}(y)+a_ip_i(y)+x_{i-1}p_{i-1}(y).
\end{equation}
The matrix $P_{ij}=p_i(\lambda^{(2)}_j)$, where $\lambda^{(2)}_j$, $j\in\{0,1,\ldots,N\}$, denotes the eigenvalues of $K_2$, defines the similarity transformation which brings the matrix $K_2$ to its diagonal form. In physical terms, given a pair  $(K_1,K_2)$ of operators expressed in the form \eqref{Matrices} acting on a state space, the polynomials defined by the recurrence relation  \eqref{recurrence-gene} are the overlap coefficients between the bases in which either $K_1$ or $K_2$ is diagonal. For more details, see \cite{Terwilliger-2001}.

\section{The Clebsch-Gordan problem}
The Clebsch-Gordan (CG) problem of $sl_{-1}(2)$ has been solved in \cite{Vinet-2011}. We recall here some of the results concerning this problem which shall prove useful. 

The CG problem can be posited in the following way. We consider the $sl_{-1}(2)$-module $(\epsilon_1,\mu_1)\otimes(\epsilon_2,\mu_2)$ or equivalently the addition of two $sl_{-1}(2)$ algebras. It is seen that the operator $J_0^{(3)}=J_0^{(1)}+J_{0}^{(2)}$ has eigenvalues of the form $\mu_1+\mu_2+N+1$, $N\in\mathbb{N}$. We denote by $\ket{q_{12},N}$ the state with eigenvalue $q_{12}$ of the Casimir operator $\mathcal{Q}_{12}$ and with a given value $N$ of  the total projection. We have
\begin{align}
\label{dompe-2}
\mathcal{Q}_{12}\ket{q_{12},N}=q_{12}\ket{q_{12},N}, && J_0^{(3)}\ket{q_{12},N}=(\mu_1+\mu_2+1+N)\ket{q_{12},N}.
\end{align}
 In view of the formula \eqref{Casimir-action}, the eigenvalues $q_{12}$ of the Casimir operator $\mathcal{Q}_{12}$ can be decomposed as the product
\begin{align}
q_{12}=-\epsilon_{12}\mu_{12},\;\;\;\;\;\epsilon_{12}=\pm 1,\;\;\mu_{12}\geqslant0,
\end{align}
whence we have $|q_{12}|=\mu_{12}$. The  Casimir operator \eqref{Casimir-added} of the added algebras can be re-expressed in terms of the local Casimir operators $\mathcal{Q}_1$ and $\mathcal{Q}_2$ in the following way:
\begin{equation}
\mathcal{Q}_{12}=\left(J_{-}^{(1)}J_{+}^{(2)}-J_{+}^{(1)}J_{-}^{(2)}\right)R^{(1)}-(1/2)R^{(1)}R^{(2)}+\mathcal{Q}_{1}R^{(2)}+\mathcal{Q}_{2}R^{(1)}.
\end{equation}
The state $\ket{q_{12},N}$ can be decomposed as a linear combination of the tensor product states
\begin{equation}
\ket{q_{12},N}=\sum_{n_1+n_2=N}C^{\mu_1\mu_2q_{12}}_{n_1n_2N}\,\ket{\epsilon_1,\mu_1,n_1}\otimes\ket{\epsilon_2,\mu_2,n_2}.
\end{equation}
The coefficients $C^{\mu_1\mu_2q_{12}}_{n_1n_2N}$ are the Clebsch-Gordan coefficients of the $sl_{-1}(2)$ algebra.
We note that these coefficients vanish unless $n_1+n_2=N$ and that their dependence on $\epsilon_1$ and $\epsilon_2$ is implicit.

The possible values of the eigenvalues $q_{12}$ of the Casimir operator $\mathcal{Q}_{12}$ are given by
\begin{equation}
\label{Spectrum-Casimir}
q_{12}=(-1)^{s+1}\epsilon_1\epsilon_2(\mu_1+\mu_2+1/2+s),\;\;\;\;\;s=0,1,\ldots,N.
\end{equation}
This result can be derived in the following way. In a given module $(\epsilon,\mu)$, the eigenvalues $\lambda_{J_0}$ of $J_0$ are 
\begin{equation}
\lambda_{J_0}=n-\epsilon \mathcal{Q}+1/2,\;\;\;\;\;n\in\mathbb{N}.
\end{equation}
 Hence, for a given eigenvalue $\lambda_{J_0}>0$ of $J_0$, the eigenvalues $q$ of the Casimir operator $\mathcal{Q}$ which are compatible with $\lambda_{J_0}$ are, in absolute value, 
\begin{equation}
|q|=|\lambda-1/2|,\,|\lambda-3/2|,\ldots
\end{equation}
When considering the coproduct of two $sl_{-1}(2)$ algebras, the eigenvalues of $J_0^{(3)}$ are $\lambda^{(3)}=\mu_1+\mu_2+N+1$. Consequently, for a given value of $N$, the set of allowed values for the eigenvalues $q_{12}$ of the Casimir $\mathcal{Q}_{12}$, which should be of cardinality $N+1$, is 
\begin{equation}
\label{range}
|q_{12}|=\mu_1+\mu_2+N+1/2,\,\mu_1+\mu_2+N-1/2,\ldots,\mu_1+\mu_2+1/2.
\end{equation}
Thus the admissible values of $\mu_{12}=|q_{12}|$ are given by the above set \eqref{range}. 

There remains to evaluate the corresponding values of $\epsilon_{12}$. To that end, we consider the eigenstate $\ket{x}$ of $\mathcal{Q}_{12}$ corresponding to the maximal value $\mu_{12})_{\text{max}}=\mu_1+\mu_2+N+1/2$. It is seen that this state satisfies the properties
\begin{equation}
\label{test}
J_0^{(3)}\ket{x}=(\mu_1+\mu_2+N+1)\ket{x}, \;\;\;\;\;J_{-}^{(3)}\ket{x}=0.
\end{equation}
On the one hand, it then follows from \eqref{test} and \eqref{action} that
\begin{equation}
R^{(3)}\ket{x}=\epsilon_{12})_{\text{max}}\ket{x},
\end{equation}
where $\epsilon_{12})_{\text{max}}$ is the value of $\epsilon_{12}$ corresponding to the maximal value of $\mu_{12}$. On the other hand, it stems from \eqref{dompe-2} and \eqref{action} that
\begin{equation}
R^{(3)}\ket{q_{12},N}=(-1)^{N}\epsilon_1\epsilon_2\ket{q_{12},N}.
\end{equation}
We thus have $\epsilon_{12})_{\text{max}}=(-1)^{N}\epsilon_1\epsilon_2$. It follows that the eigenvalue $q_{12}$ of the Casimir operator $\mathcal{Q}_{12}$ corresponding to the maximal value of $|q_{12}|$ is given by
\begin{equation}
q_{12}=(-1)^{N+1}\epsilon_1\epsilon_2(\mu_1+\mu_2+1/2+N).
\end{equation}
By induction on $N$, one is led to the announced form of the eigenvalues \eqref{Spectrum-Casimir}. The Casimir operator $\mathcal{Q}_{12}$ is tridiagonal in the tensor product basis. This allows to obtain a recurrence relation for the Clebsch-Gordan coefficients which, given the spectrum \eqref{Spectrum-Casimir}, is seen to coincide with that of the dual $-1$ Hahn polynomials \cite{VZhedanov-2011,Vinet-2011}.                                                                                                                                                                                                                                                                                                                                                                                                                                                                                                                                                                                                                                                                                                                                                                                                                                                                                   
\section{The Racah problem and Bannai-Ito algebra}
The addition rule \eqref{addition-rule} possess an associativity property when three $sl_{-1}(2)$ algebras are added. We consider three mutually commuting sets of $sl_{-1}(2)$ generators $\scriptstyle J_0^{(j)}$, $\scriptstyle J_{\pm}^{(j)}$ and $\scriptstyle R^{(j)}$ for $j=1,2,3$. The resulting fourth algebra can be obtained by two different addition sequences.  Indeed, one has the two equivalent schemes: $4=(1\oplus 2)\oplus 3$ and $4=1\oplus(2\oplus 3)$. The Racah problem consists in finding the overlap between the respective eigenstates of the intermediary Casimir operators $\mathcal{Q}_{12}$ and $\mathcal{Q}_{23}$ with a fixed eigenvalue $q_4$ of the total Casimir operator $\mathcal{Q}_4$. Denoting such eigenstates by $\ket{q_{12};q_4,m}$ and $\ket{q_{23};q_4,m}$, the Racah coefficients are defined as
\begin{equation}
\ket{q_{12};q_4,m}=\sum_{q_{23}}R^{\mu_1\mu_2\mu_3}_{q_{12}q_{23}\mu_4}\ket{q_{23};q_4,m},
\end{equation}
where we have by definition
\begin{align}
\label{eigen}
\mathcal{Q}_{12}\ket{q_{12},q_4,m}=q_{12}\ket{q_{12},q_4,m},\;\;\;\;\; \mathcal{Q}_{23}\ket{q_{23},q_4,m}=q_{23}\ket{q_{23},q_4,m}.
\end{align}
We note that the Racah coefficients $R^{\mu_1\mu_2\mu_3}_{q_{12}q_{23}\mu_4}$ do not depend on the total projection number $m$ and that their dependence on $\epsilon_i$, $i\in\{1,\ldots,4\}$, is implicit. The problem of finding the overlap coefficients is non-trivial because the operators $\mathcal{Q}_{12}$ and $\mathcal{Q}_{23}$ do not commute, hence they cannot be simultaneously diagonalized. The two intermediary Casimir operators have the following expressions:
\begin{align}
\label{ops}
K_1=\mathcal{Q}_{12}&=\left(J_{-}^{(1)}J_{+}^{(2)}-J_{+}^{(1)}J_{-}^{(2)}\right)R^{(1)}-R^{(1)}R^{(2)}/2+\mathcal{Q}_{1}R^{(2)}+\mathcal{Q}_{2}R^{(1)},\\
\label{ops2}
K_2=\mathcal{Q}_{23}&=\left(J_{-}^{(2)}J_{+}^{(3)}-J_{+}^{(2)}J_{-}^{(3)}\right)R^{(2)}-R^{(2)}R^{(3)}/2+\mathcal{Q}_{2}R^{(3)}+\mathcal{Q}_{3}R^{(2)}.
\end{align}
The full Casimir operator of the fourth algebra $\mathcal{Q}_4$ can also be obtained in a straightforward manner; one finds
\begin{equation}
\label{Full-Casimir}
\mathcal{Q}_{4}=\left(J_{-}^{(1)}J_{+}^{(3)}-J_{+}^{(1)}J_{-}^{(3)}\right)R^{(1)}-\mathcal{Q}_2R^{(1)}R^{(3)}+\mathcal{Q}_{12}R^{(3)}+\mathcal{Q}_{23}R^{(1)}.
\end{equation}
The paramount observation is that the operators $K_1$, $K_2$ are closed in frames of a simple algebra with three generators. To see this, one first defines
\begin{equation}
\label{ops3}
K_3=(J_{+}^{(1)}J_{-}^{(3)}-J_{-}^{(1)}J_{+}^{(3)})R^{(1)}R^{(2)}+R^{(1)}R^{(3)}/2-\mathcal{Q}_1R^{(3)}-\mathcal{Q}_3R^{(1)}.
\end{equation}
Since the operators $\mathcal{Q}_i$ for $i=1,\ldots,4$ commute with $K_1$, $K_2$, $K_3$ and among themselves, we shall replace them by their corresponding eigenvalues $-\lambda_j$ where $\lambda_j=\epsilon_j\mu_j$. A direct computation shows that the following relations hold:
\begin{align}
\label{BI-Algebra}
\{K_1,K_{2}\}=K_{3}+\alpha_3, && \{K_2,K_3\}=K_1+\alpha_1, && \{K_1,K_3\}=K_2+\alpha_2,
\end{align}
where the structure constants are given by
\begin{align}
\label{StructureConstants}
\alpha_{1}=-2(\lambda_1\lambda_2+\lambda_3\lambda_4), && \alpha_{2}=-2(\lambda_1\lambda_4+\lambda_2\lambda_3), && \alpha_3=2(\lambda_1\lambda_3+\lambda_2\lambda_4).
\end{align}
Note that the first relation of \eqref{BI-Algebra} can be considered as a \emph{definition} of $K_3$. The algebra \eqref{BI-Algebra} is known as the Bannai-Ito algebra \cite{Vinet-2012}, which is, as will be seen below, a special case of the Askey-Wilson algebra \eqref{AW-relation}. It admits the Casimir operator
\begin{equation}
\mathcal{Q}_{BI}=K_1^2+K_2^2+K_3^2,
\end{equation}
which commutes with all generators. Given the realization \eqref{StructureConstants} of this algebra, the Casimir operator takes the value
\begin{equation}
\mathcal{Q}_{BI}=\lambda_1^2+\lambda_2^2+\lambda_3^2+\lambda_4^2-1/4.
\end{equation}
We now look to construct irreducible BI-modules;  the degree of these representations is prescribed by the range of possible eigenvalues of the operators $\mathcal{Q}_{12}$ and $\mathcal{Q}_{23}$. For simplicity, we restrict ourselves to the case where $\epsilon_1$, $\epsilon_2$ and $\epsilon_3$ are all equal to $1$. The other cases can be treated in similar fashion. It is worth mentioning that $\epsilon_4$ cannot be fixed \emph{a priori} and will in fact depend on the degree of the given module. 

In the CG problem, the possible eigenvalues of $q_{12}$ were determined by the value of the total projection operator. For the Racah problem, the overlap coefficients are independent of the total projection and the spectrum of $\mathcal{Q}_{12}$ is restricted only by the value of the total Casimir operator $\mathcal{Q}_4$. From \eqref{range}, one finds that the minimal value of the absolute value of $q_{12}$ is given by
\begin{equation}
\label{min}
|q_{12}|_{\text{min}}=\mu_1+\mu_2+1/2.
\end{equation}
In addition, in view of the addition scheme $4=(1\oplus2)\oplus 3$, we have that the absolute value of the eigenvalues $q_{4}$ of the total Casimir operator $\mathcal{Q}_4$ are of the form
\begin{equation}
|q_4|=|q_{12}|+\mu_3+1/2+s_{12,3},\;\;\;\;\;s_{12,3}=0,1,\ldots,m
\end{equation}
where $m$ is the total projection of the state $\ket{q_{12};q_4,m}$. It is clear that for a given absolute value of $|q_4|=\mu_4$, the maximal value of $|q_{12}|$ corresponds to setting $s_{12,3}=0$. It then follows that
\begin{equation}
\label{max}
|q_{12}|_{\text{max}}=\mu_4-\mu_3-1/2.
\end{equation}
Considering finite-dimensional representations of degree $N+1$, we find from \eqref{min} that the eigenvalues $q_{12}$ of the Casimir $\mathcal{Q}_{12}$ are of the form
\begin{equation}
\label{range-2}
|q_{12}|=\mu_1+\mu_2+1/2,\mu_1+\mu_2+3/2,\ldots, \mu_1+\mu_2+1/2+N.
\end{equation}
Using \eqref{max} and \eqref{range-2}, we obtain
\begin{equation}
\label{degree}
N+1=\mu_4-\mu_1-\mu_2-\mu_3.
\end{equation}
The spectra of the intermediary Casimir operators $\mathcal{Q}_{12}$ and $\mathcal{Q}_{23}$ are thus given by
\begin{align}
\label{eigenvalues-2}
q_{12}&=(-1)^{s_{12}+1}(\mu_1+\mu_2+1/2+s_{12}),\;\;\;\;\; s_{12}=0,\ldots,N, \\
q_{23}&=(-1)^{s_{23}+1}(\mu_2+\mu_3+1/2+s_{23}),\;\;\;\;\; s_{23}=0,\ldots,N.
\end{align}
The parameter $\epsilon_4$ is prescribed by the value of $N$. Indeed, from the CG problem it is known that the allowed eigenvalues $q_{4}$ of the Casimir operator $\mathcal{Q}_4$ are of the form
$$
q_4=(-1)^{k+1}(\mu_{12}+\mu_3+1/2+k),\;\;\;\;\;k=0,\ldots,m
$$
where $m$ is the total projection. Taking into account the condition \eqref{degree}, one finds
\begin{equation}
\epsilon_4=(-1)^N.
\end{equation}
Having found the explicit expressions for the spectra and dimension in terms of the representation parameters, the matrix representation of the BI algebra can be made explicit.
\section{Leonard pair and Racah coefficients}
Let $\mu_1$, $\mu_2$, $\mu_3$ be fixed (positive) representation parameters and $N$ a positive integer as in \eqref{degree}. The operators $K_1$, $K_2$ are  square matrices of dimension $N+1$ which are easily seen to satisfy the following Askey-Wilson relations:
\begin{align}
\label{AW-1}
K_1^2K_2+2K_1K_2K_1+K_2K_1^2-K_2=\kappa_3K_1+\kappa_2,\\
\label{AW-2}
K_2^2K_1+2K_2K_1K_2+K_1K_2^2-K_1=\kappa_3K_2+\kappa_1,
\end{align}
where the constants $\kappa_i$, $i=1,2,3$, are given by
\begin{align}
\begin{split}
\kappa_1&=-2(\mu_1\mu_2+\epsilon_4\mu_3\mu_4),\\
\kappa_2&=-2(\mu_2\mu_3+\epsilon_4\mu_1\mu_4), \\
 \kappa_3&=4(\mu_1\mu_3+\epsilon_4\mu_2\mu_4),
\end{split}
\end{align}
with $\epsilon_4=(-1)^{N}$. The matrix representing $K_1$ can be made diagonal with eigenvalues prescribed by \eqref{eigenvalues-2} and \eqref{eigen}. In this basis, it is easily seen that the relations \eqref{AW-1} and \eqref{AW-2} imply that $K_2$ must be irreducible tridiagonal. The pair $(K_1,K_2)$ thus forms a Leonard pair. Consequently, there exists a basis in which the matrices $K_1$ and $K_2$ can be expressed as
\begin{align}
K_1=\mathrm{diag}\,(\theta_0,\theta_1,\cdots,\theta_N), && K_2=
\begin{pmatrix}
b_0 & 1 &   &  & & \mathbf{0}\\
u_1 & b_1 & 1 & & \\
 & u_2 & b_2 & 1 & & \\
 & & \ddots & \ddots & \ddots & \\
 & & & & b_{N-1}& 1 \\
\mathbf{0} & &&&u_N& b_N                                                 
 \end{pmatrix},
\end{align}
where $\theta_i=(-1)^{i+1}(\mu_1+\mu_2+1/2+i)$ for $i=0,\ldots,N$ and $u_n$, $b_n$ are indeterminate. Imposing the relations \eqref{AW-1} and \eqref{AW-2} on the two operators and solving for $b_n$ and $u_n$, one finds
\begin{align}
\label{Central}
-b_n=
\begin{cases}
(\mu_2+\mu_3+1/2)+\frac{1}{2}\frac{n(n+\mu_1+\mu_2-\mu_3-\epsilon_4\mu_4)}{(n+\mu_1+\mu_2)}\\
-\frac{1}{2}\frac{(n+1+2\mu_2)(n+1+\mu_1+\mu_2+\mu_3-\epsilon_4\mu_4)}{(n+1+\mu_1+\mu_2)}, & \text{for $n $ even,} \\
(\mu_2+\mu_3+1/2)+\frac{1}{2}\frac{(n+2\mu_1)(n+\mu_1+\mu_2-\mu_3+\epsilon_4\mu_4)}{(n+\mu_1+\mu_2)}\\
-\frac{1}{2}\frac{(n+1+2\mu_1+2\mu_2)(n+1+\mu_1+\mu_2+\mu_3+\epsilon_4\mu_4)}{(n+1+\mu_1+\mu_2)},
& \text{for $n$ odd,}
\end{cases}
\end{align}
\begin{align}
\label{Down}
u_n &=
\begin{cases}
-\frac{1}{4}\frac{n(n+2\mu_1+2\mu_2)(n+\mu_1+\mu_2+\mu_3+\epsilon_4\mu_4)(n+\mu_1+\mu_2-\mu_3-\epsilon_4\mu_4)}{(n+\mu_1+\mu_2)^2}, & \text{for $n$ even,} \\
-\frac{1}{4}\frac{(n+2\mu_2)(n+2\mu_1)(n+\mu_1+\mu_2+\mu_3-\epsilon_4\mu_4)(n+\mu_1+\mu_2-\mu_3+\epsilon_4\mu_4)}{(n+\mu_1+\mu_2)^2}, & \text{for $n$ odd.}
\end{cases}
\end{align}
The overlap coefficients of the bases in which either $\mathcal{Q}_{12}$ or $\mathcal{Q}_{23}$ is diagonal will thus be proportional to the monic polynomials $\tilde{P}_{n}(\theta^*_i)$ with $\theta^*_i=(-1)^{i+1}(\mu_2+\mu_3+1/2+i)$ which obey the recurrence relation
\begin{equation}
\tilde{P}_{n+1}(x)+b_n\tilde{P}_{n}(x)+u_n\tilde{P}_{n-1}(x)=x\tilde{P}_{n}(x).
\end{equation}
Defining $P_{n}(x)=(-2)^{-n}\tilde{P}_{n}(x)$, we recover the recurrence relation of the monic Bannai-Ito polynomials 
\begin{equation}
P_{n+1}(x_s)+(\rho_1-A_n-C_n)P_n(x_s)+A_{n-1}C_{n}P_{n}(x_s)=x_sP_n(x_s),
\end{equation}
with $x_s=-\theta^*_s/2-1/4$ and where the identification with the Bannai-Ito parameters is
\begin{align}
\label{identification}
\begin{split}
\rho_1&=\frac{1}{2}(\mu_2+\mu_3),\;\;\;\;\; \rho_2=\frac{1}{2}(\mu_1+\epsilon_4\mu_4), \\
r_1&=\frac{1}{2}(\mu_3-\mu_2), \;\;\;\;\; r_2=\frac{1}{2}(\epsilon_4\mu_4-\mu_1).
\end{split}
\end{align}
The coefficients $A_n$ and $C_n$ are as defined in \eqref{Coeff-1}. The truncation conditions are the following. On the one hand, if $N$ is even, we have
\begin{equation}
\label{Identif-1}
2(r_2-\rho_1)=N+1,
\end{equation}
as well as the identification $a=\mu_1$, $b=\mu_2$ and $c=\mu_3$. 
On the other hand, if $N$ is odd, we have
\begin{equation}
\label{Identif-2}
2(\rho_1+\rho_2)=-(N+1),
\end{equation}
and the identification $\alpha=\mu_1$, $\beta=\mu_2$ and $\gamma=\mu_3$. Is is seen that the Bannai-Ito grids \eqref{Grille-1} and \eqref{Grille-2} coincide, as expected, with the predicted spectrum of the Casimir operator $\mathcal{Q}_{23}$. To determine the normalization constant, we use the unitarity of the transformation which imposes the following orthogonality relation for Racah coefficients:
\begin{equation}
\label{Ortho-Racah}
\sum_{q_{23}}R^{\mu_1\mu_2\mu_3}_{q q_{23}\mu_4}R^{\mu_1\mu_2\mu_3}_{q'q_{23}\mu_4}=\delta_{qq'}.
\end{equation}
Using the relations \eqref{Ortho-Pair}, \eqref{Ortho-Odd}  and \eqref{Ortho-Racah}, we obtain
\begin{equation}
\label{Racah}
R^{\mu_1\mu_2\mu_3}_{q_{12}q_{23}\mu_4}=\sqrt{\frac{\Omega_{\ell}(x_{\ell})}{\Phi_{N,n}}}P_{n}(\rho_1,\rho_2,r_1,r_2; x_{\ell}),
\end{equation}
where $P_n(\rho_1,\rho_2,r_1,r_2; x_{\ell})$ is the monic Bannai-Ito polynomial. In addition, we have
\begin{align}
x_{\ell}=-\frac{1}{2}\left(\theta^*_{\ell}+1/2 \right),&& \ell=|q_{23}|-\mu_2-\mu_3-1/2, &&& n=|q_{12}|-\mu_1-\mu_2-1/2,
\end{align}
along with the identifications \eqref{degree}, \eqref{identification}, \eqref{Identif-1} and \eqref{Identif-2}. The Racah coefficients \eqref{Racah} are thus determined up to a phase factor. Returning the Bannai-Ito algebra \eqref{BI-Algebra}, it is seen that the realization \eqref{StructureConstants} is invariant under the permutations $\pi_1=(12)(34)$, $\pi_2=(13)(24)$ and $\pi_3=(14)(23)$ of the representation parameters $\lambda_i$, $i=1,\ldots4$. These transformations generate the Klein four-group. In addition, the operation $\lambda_i\rightarrow -\lambda_i$ also leaves \eqref{BI-Algebra} and \eqref{StructureConstants} invariant.
\section{The Racah problem for the addition of ordinary oscillators}
When $\mu=0$, the $sl_{-1}(2)$ algebra reduces to the Heisenberg oscillator algebra endowed however with a non-trivial coproduct. Therefore, the algebra obtained from the Hopf addition rule \eqref{addition-rule} of two $sl_{-1}(2)$ algebras with $\mu_i=0$ is not as a result a pure oscillator algebra, but a parabosonic algebra. The same assertion holds for the addition of three $sl_{-1}(2)$ algebras with Casimir parameters $\mu_1$, $\mu_2$ and $\mu_3$ all equal to zero. This corresponds to adding three pure oscillator algebras with the addition rule \eqref{addition-rule}. Due to the importance of the oscillator algebra, it is worth recording this reduction in some detail. Most algebraic results connected to this skewed addition of three quantum harmonic oscillators have interestingly been obtained previously in \cite{Gorodnii-1984,Silvestrov-1992,Arik-2003,Brown-2012}. In the case $\mu_1=\mu_2=\mu_3=0$, the Bannai-Ito \eqref{BI-Algebra} algebra becomes
\begin{align}
\label{Anti-su}
\{K_1,K_2\}=K_3, && \{K_2,K_3\}=K_1, &&& \{K_1,K_3\}=K_2.
\end{align}
This algebra can be seen as an anti-commutator version of the classical $\mathfrak{su}(2)$ Lie algebra. The Askey-Wilson relations simplify to
\begin{align}
\label{AW-3}
K_1^2K_2+2K_1K_2K_1+K_2K_1^2-K_2=0,\\
\label{AW-4}
K_2^2K_1+2K_2K_1K_2+K_1K_2^2-K_1=0.
\end{align}
The spectra of the operators $K_1$ and $K_2$ are then given by the formula
\begin{align}
\theta_{i}=(-1)^{i+1}(i+1/2).
\end{align}
Moreover, the degree of the module is $N+1=\mu_4$ with $\epsilon_4=(-1)^{N}$. With these observations, the pair $(K_1,K_2)$ again forms a Leonard pair. The matrices $K_1$ and $K_2$ can thus be put in the form
\begin{align}
K_1=\mathrm{diag}\,(\theta_0,\theta_1,\cdots,\theta_N), && K_2=
\begin{pmatrix}
b_0 & 1 &   &  & & \mathbf{0}\\
u_1 & b_1 & 1 & & \\
 & u_2 & b_2 & 1 & & \\
 & & \ddots & \ddots & \ddots & \\
 & & & & b_{N-1}& 1 \\
\mathbf{0} & &&&u_N& b_N                                                 
 \end{pmatrix}.
\end{align}
In this case, solving for the coefficients $b_n$ and $u_n$ yields on the one hand
\begin{equation}
b_0=-(N+1)/2,\;\;\;\;\;\;\;\text{and }\;\;\;\;\;\;\; b_i=0\text{ for $i\neq 0$},
\end{equation}
and on the other hand
\begin{equation}
u_n=\frac{(n+N+1)(N+1-n)}{4}.
\end{equation}
The positivity and truncation conditions $u_n>0$ and $u_{N+1}$ are manifestly satisfied here. As expected, the obtained sequences $\{b_n\}$, $\{u_n\}$ correspond to the specializations $\mu_1=\mu_2=\mu_3=0$ of the formulas \eqref{Central} and \eqref{Down}. Similarly to the Bannai-Ito case, the similarity transformation bringing $K_2$ into its diagonal form can be constructed with the Bannai-Ito polynomials reduced with the parametrizations $a=0$, $b=0$ and $c=0$ in the $N$ even case and $\alpha=0$, $\beta=0$ and $\gamma=0$ in the $N$ odd case. The explicit hypergeometric representation \eqref{Hypergeo-representation} of the corresponding polynomials, the weight functions \eqref{weight-even}, \eqref{weight-odd} as well as the normalization constants can be imported directly without need of a  limiting procedure.
\phantom{\cite{Zhedanov-1993}}
\section*{Conclusion: the Leonard triple}
We considered the Racah problem for the algebra $sl_{-1}(2)$ which acts as the dynamical algebra for a \emph{parabosonic oscillator} and showed that the algebra of the intermediary Casimir operators coincide with the Bannai-Ito algebra. From the knowledge of the Clebsch-Gordan problem, the spectra of the Casimir operators were determined and this allowed to build the relevant finite-dimensional modules for the BI algebra. It was then recognized that the operators $\mathcal{Q}_{12}=K_1$ and $\mathcal{Q}_{23}=K_2$ form a Leonard pair and this observation was used to see that the overlap (Racah) coefficients are given in terms of the Bannai-Ito polynomials. 

As is manifest from \eqref{BI-Algebra}, the Bannai-Ito algebra has a $Z_3$ symmetry with respect to a relabeling of the operators $K_i$ with $i=1,2,3$. However, the Racah problem considered here provides a specific realization of the BI algebra in terms of the distinct operators $\mathcal{Q}_{12}$, $\mathcal{Q}_{23}$ and $K_3$, for which this symmetry is not present. In this regard, it is natural to ask whether there exists a situation for which it is the pair $(K_2,K_3)$ or $(K_1,K_3)$ that is realized by intermediate Casimir operators. This question can answered by considering the Racah problem for the addition of three $sl_{-1}(2)$ algebras with different addition rules that lead to a fourth algebra that has nevertheless the same total Casimir $\mathcal{Q}_{4}$. The first intermediate algebra $\widetilde{(31)}=\widetilde{3}\oplus \widetilde{1}$ is obtained by defining
\begin{align}
\label{voila}
J_0^{(31)}=J_0^{(1)}+J_0^{(3)}, && J_{\pm}^{(31)}=J_{\pm}^{(1)}R^{(3)}+J_{\pm}^{(3)}R^{(2)}, &&& R^{(31)}=R^{(1)}R^{(3)},
\end{align}
which differs from the original coproduct by the presence of $R^{(2)}$. Note that \eqref{voila} implicitly uses $(\epsilon_2,\mu_2)$ as an auxiliary space. The intermediate Casimir operator $\widetilde{\mathcal{Q}}_{13}$ is then found to coincide with the negative of $K_3$ as defined in \eqref{ops3}:
$$
\widetilde{\mathcal{Q}}_{31}=-K_3.
$$
A second intermediate Casimir operator is obtained by using the standard coproduct \eqref{addition-rule} in two ways: one forms the algebra $(12)$, for which $\widetilde{Q}_{12}=K_1$, or one forms the algebra $(23)$ for which $\widetilde{Q}_{23}=K_2$. To ensure consistency, as mentioned before, the full Casimir operator of the fourth algebra $(4)=\widetilde{(31)}\oplus \widetilde{(2)}$ should coincide with \eqref{Full-Casimir}. This is done by defining
\begin{align}
\label{algebre-4}
J_0^{(4)}=J_0^{(31)}+J_0^{(2)}, && J_{\pm}^{(4)}=J_{\pm}^{(31)}R^{(2)}+J_{\pm}^{(2)}R^{(3)}, &&& R^{(4)}=R^{(31)}R^{(2)},
\end{align}
It is readily seen that the generators defined in \eqref{voila} and \eqref{algebre-4} satisfy the defining relations \eqref{Definition} of $sl_{-1}(2)$. This fourth algebra is easily seen to admit the same full Casimir operator \eqref{Full-Casimir}. Defining $\widetilde{K_3}=-\widetilde{\mathcal{Q}}_{31}$, $\widetilde{K_1}=K_1$ and $\widetilde{K_2}=K_2$, the algebra \eqref{BI-Algebra} is recovered with the pair $(\widetilde{K_1},\widetilde{K_3})$ or $(\widetilde{K_2},\widetilde{K_3})$ playing the role of the intermediate Casimir operators. The steps of Sections 3, 4 can then be reproduced and this leads one to conclude that $K_3$ also has a Bannai-Ito type spectrum $\lambda_{i}^{(3)}=(-1)^{i}(\mu_1+\mu_3+1/2+i)$, $i=0,\ldots,N$ and that $(K_2,K_3)$ and $(K_1,K_3)$ form Leonard pairs. In addition, it follows from this observation that in the realization \eqref{StructureConstants} of the Bannai-Ito algebra \eqref{BI-Algebra} obtained from the operators \eqref{ops}, \eqref{ops2} and \eqref{ops3}, the set $(K_1,K_2,K_3)$ constitutes a \emph{Leonard Triple}, which have studied intensively for the $q$-Racah scheme in \cite{Curtin-2007,Huang-2011}.

In the case of the algebras $\mathfrak{sl}(2)$ and $sl_{q}(2)$, it is known that the Clebsch-Gordan coefficients can be obtained from the Racah coefficients in a proper limit. It is not so with the algebra $sl_{-1}(2)$. Indeed, the dual $-1$ Hahn polynomials are beyond the Leonard duality and do not occur as limits of the Bannai-Ito polynomials. Furthermore, the question of the symmetry algebra underlying the Clebsch-Gordan problem for $sl_{-1}(2)$ remains open. We plan to report on this elsewhere.

\section*{Acknowledgments}
\noindent
V.X.G. holds a scholarship from Fonds de recherche du Qu\'ebec-Nature et technologies (FQRNT). The research of L.V. is supported in part by the Natural Sciences and Engineering Research Council (NSERC) of Canada. A.Z. wishes to thank the Centre de Recherches Math\'ematiques (CRM) for the hospitality extended to him.

\bibliographystyle{amsplain}


\end{document}